\documentclass{ws-procs9x6}

\newcommand{\gae}{$\stackrel{>}{\sim}$}
\newcommand{\laem}{\stackrel{<}{\sim}}
\newcommand{\gaem}{\stackrel{>}{\sim}}
\newcommand{\beq}{\begin{equation}}
\newcommand{\eeq}{\end{equation}}
\newcommand{\beqa}{\begin{eqnarray}}
\newcommand{\eeqa}{\end{eqnarray}}

\newcommand{\NPB}[1]{{\it Nucl. Phys.}\ {\bf B{#1}}}
\newcommand{\PLB}[1]{{\it Phys. Lett.}\ {\bf B{#1}}}
\newcommand{\PRD}[1]{{\it Phys. Rev.}\ {\bf D{#1}}}

\begin{document}

\title{Flavor Constraints on Theory Space}

\author{E.~H. SIMMONS and R.~S. CHIVUKULA}

\address{Department of Physics, Boston University, \\
590 Commonwealth Avenue, \\ 
Boston, MA 02215, USA\\ 
E-mail: simmons@bu.edu, sekhar@bu.edu}

\author{N. EVANS}

\address{Department of Physics, University of Southampton, \\ 
Highfield, Southampton, SO17 1BJ, UK \\
E-mail: evans@phys.soton.ac.uk}  


\maketitle

\abstracts{Composite Higgs models based on the chiral symmetries of
``theory space'' can produce Higgs bosons with masses of order 100 GeV
from underlying strong dynamics at scales up to 10 TeV without fine
tuning.  This talk argues that flavor-violating interactions
generically arising from underlying flavor dynamics constrain the
Higgs compositeness scale to be \gae 75 GeV, implying that significant
fine-tuning is required. Bounds from CP violation and weak isospin
violation are also discussed. }

\section{Introduction}

The Standard Higgs Model employs a fundamental scalar doublet to break
the electroweak symmetry and provide fermion masses.  Well-known
difficulties, including the hierarchy problem and the
triviality problem, imply that the Standard Higgs Model is just a
low-energy effective theory.

Suppose that the Higgs field $\phi$ is actually a
composite\cite{chiggs} state arising from underlying strong dynamics
at a higher energy scale, $\Lambda$.  We can estimate the sizes of
operators involving $\phi$ in the low-energy effective theory using
dimensional analysis\cite{ndaref}. A theory with light scalar
particles in a single symmetry-group representation
depends\cite{generalized} on two parameters: $\Lambda$, the scale of
the underlying physics, and $f$, the analog of $f_\pi$ in QCD. Our
estimates of the low-energy effects of the underlying
physics will depend on $\kappa \equiv \Lambda / f$.

Regardless of the precise nature of the underlying
strongly-interacting physics that produces $\phi$, there must be
flavor dynamics at a scale $\gaem\Lambda$ that gives rise to the
different Yukawa couplings of the Higgs boson to ordinary fermions.
If this flavor dynamics arises from gauge interactions it will
generally cause flavor-changing neutral currents (as in ETC models
\cite{etc}).  Similarly, there are likely to be couplings that violate
CP and weak isospin.  This talk reviews the
constraints\cite{Chivukula:2002ww} which FCNC, CP-violation, and weak-isospin violation place
on composite Higgs models and applies the limits to models developed
\cite{Arkani-Hamed:2001nc,Arkani-Hamed:2002pa} under the rubric of
``theory space''\cite{Arkani-Hamed:2001ca}.

\section{Composite Higgs Phenomenology}
\label{sec:flavor}
\subsection{Flavor}

Quark Yukawa couplings arise from flavor physics coupling the
left-handed doublets $\psi_L$ and right-handed singlets $q_R$ to the
strongly-interacting constituents of the composite Higgs doublet.  If
these new flavor interactions are gauge interactions with gauge
coupling $g$ and gauge boson mass $M$, dimensional analysis
\cite{ndaref} estimates the resulting Yukawa coupling is \cite{chiggs}
of order ${g^2\over M^2} {\Lambda^2 \over \kappa}$.  To produce a
quark mass $m_q$, the Yukawa coupling must equal ${\sqrt{2} m_q / v}$
where $v\approx 246$ GeV. This implies\cite{Chivukula:1996rz}
\beq
\Lambda \gaem {M \over g} \sqrt{\sqrt{2} \kappa {m_q \over v}}~.
\label{eq:yukawa}
\eeq 
If experiment sets a lower limit on $M/g$,
eqn.(\ref{eq:yukawa}) gives a lower bound on $\Lambda$.

Consider the interactions responsible
for the $c$-quark mass. Through Cabibbo mixing, these interactions must
couple to the $u$-quark as well:
\beqa
{L}_{eff} & = & - \, (\cos \theta_L^c \sin \theta_L^c)^2 
\frac{g^{2}}{M^{2}}
( \overline c_L  \gamma^{\mu} u_L )(\overline c_L  \gamma_{\mu} u_L)
\nonumber \\ [2mm]
& & -  \, (\cos \theta_R^c \sin \theta_R^c)^2 
\frac{g^{2}}{M^{2}}
( \overline c_R \gamma^{\mu} u_R)(\overline c_R \gamma_{\mu} u_R)
\nonumber \\ [2mm]
& & - \,
2 \cos \theta_L^c \sin \theta_L^c \cos \theta_R^c \sin \theta_R^c
\frac{g^{2}}{M^{2}}
( \overline c_L  \gamma^{\mu} u_L )(\overline c_R \gamma_{\mu} u_R)~,
\label{ops1}
\eeqa
where $g$ and $M$ are of the same order as those which produce
the $c$-quark Yukawa
coupling, and $\theta^c_L$, $\theta^c_R$ relate the gauge 
and  mass eigenstates.  

The color-singlet products of currents in eqn.~(\ref{ops1}) 
contribute to $D$-meson mixing.  The left-handed or
right-handed current-current operators yield\cite{Chivukula:1996rz}
\beq
\left(\frac{M}{g}\right)_{ \! {\rm LL,RR}} 
\gaem 
f_D\left( \frac{2  m_D B_D}{3 \Delta m_D }\right)^{\! 1/2}
\cos \theta_{L,R}^c \sin \theta_{L,R}^c \approx 225 \, {\rm TeV} ~,
\eeq
where
$\Delta m_D \laem  4.6 \times  10^{-11}$ MeV \cite{PDG},
and $f_D \sqrt{B_D} = 0.2$ GeV \cite{FDREF}, $\theta_{L,R}^c \approx \theta_C$.
A bound\cite{Chivukula:2002ww} on the scale of the underlying 
dynamics follows from eqn.~(\ref{eq:yukawa}): 
\beq
\Lambda \gaem 21 \, {\rm  TeV}
\sqrt{\kappa\left({m_c\over 1.5\, {\rm GeV}}\right)}~,
\label{eq:Dbound}
\eeq
so that $\Lambda \gaem 74$ TeV for $\kappa \approx 4\pi$.
The LR product of color-singlet currents gives a
weaker bound than eqn.~(\ref{eq:Dbound}).
The LR product of
color-octet currents,
\beq
{L}_{eff} = - \,
2 \cos \theta_L^c \sin \theta_L^c \cos \theta_R^c \sin \theta_R^c
\frac{g^2}{M^2}
( \overline c_L \gamma^{\mu} T^a u_L)
(\overline c_R \gamma_{\mu} T^a u_R) ~,
\label{ops2}
\eeq
where $T^a$ are the generators of $SU(3)_C$, gives a stronger bound\cite{Chivukula:2002ww}:
\beq
\Lambda \gaem 53 \, {\rm  TeV}
\sqrt{\kappa\left({1.5\, {\rm GeV}\over m_c}\right)}~.
\label{eq:DDbound}
\eeq

Analogous bounds on $\Lambda$ can be derived from neutral Kaon mixing.  However, because $m_s \ll m_c$, while the $d-s$ and $u-c$ mixings are
expected to be of comparable size, these bounds on $\Lambda$ are weaker than (\ref{eq:Dbound})
\cite{Chivukula:1997iw}. 

\subsection{Isospin}

Weak-isospin violation is a key issue in composite Higgs models
\cite{Chivukula:1997iw,Chivukula:1996sn,Chivukula:1996rz,Chivukula:1999az}.
The standard one-doublet Higgs model has an accidental custodial
isospin symmetry \cite{custodial}, which implies $\rho \approx 1$.
While all operators of 
dimension $\leq 4$ automatically respect custodial
symmetry, terms of higher dimension that arise from the underlying
physics at scale $\Lambda$ in general will not.

The leading custodial-symmetry violating operator
\beq
{\kappa^2 \over \Lambda^2} 
(\phi^\dagger D^\mu \phi)
(\phi^\dagger D_\mu \phi)~
\eeq
gives rise to a contribution to the $\rho$ parameter\cite{Chivukula:1996sn}
\beq
\Delta \rho * = - O(\kappa^2 \frac{v^2}{\Lambda^2})~.
\eeq
The limit $\vert \Delta\rho *\vert \laem 0.4\% $ implies 
$\Lambda \gaem 4 {\rm TeV} \cdot \kappa$.

\subsection{CP Violation}

In the absence of additional
superweak interactions to give rise to CP-violation in $K$-mixing
($\varepsilon$), the flavor interactions responsible for the $s$-quark
Yukawa couplings must do so. This yields strong
bounds on $\Lambda$.  Recalling 
\beq
 {\rm Re}\, \varepsilon \approx  { {\rm Im M_{12}} \over {2\, \Delta M}} 
 \laem 1.65\,\times\,10^{-3}\, ,
\eeq
and assuming that there are phases of order 1 in the $\Delta S=2$
operators analogous to those shown in eqn. (\ref{ops1}), we find 
\beq
\Lambda \gaem 120\, {\rm TeV} \sqrt{\kappa \left({m_s \over 200\, 
{\rm MeV}}\right)}~.
\label{eq:Ebound}
\eeq

\section{Composite Higgs Bosons from Theory Space}


A set of ``theory space'' composite Higgs models
\cite{Arkani-Hamed:2001nc,Arkani-Hamed:2002pa} can be represented as
an $N\times N$ toroidal lattice of ``sites'' connected by ``links'',
using ``moose'' or ``quiver'' notation \cite{moose}.  Each site except
$(1,1)$ represents a gauged $SU(3)$ group, while the links represent
non-linear sigma fields transforming as $(N,\bar{N})$'s under the
adjacent groups.  At the site $(1,1)$, only the $SU(2)\times U(1)$
subgroup of an $SU(3)$ global symmetry is gauged.  For simplicity, we
will assume the gauge couplings of the $SU(3)$ gauge groups are the
same for every site (except $(1,1)$).  Calling the ``pion decay
constant'' of the chiral-symmetry-breaking dynamics $f$, dimensional
analysis \cite{ndaref} then implies that the scale $\Lambda$ of the
underlying high-energy dynamics which gives rise to this theory is
$\laem 4\pi f$.

The $2N^2$ Goldstone bosons of the chiral symmetry breaking dynamics
are incorporated into the sigma-model fields.  As described in
\cite{Arkani-Hamed:2001nc,Arkani-Hamed:2002pa}, $N^2-1$ sets of
Goldstone bosons are eaten, $N^2-1$ get mass from ``plaquette
operators'' which explicitly break the chiral symmetries, and two sets
which are uniform in the `u' or `v' directions, along the lattice axes, remain massless in the very low-energy theory:
Both the $\pi_u$ and $\pi_v$ fields contain $SU(2)\times U(1)$ doublet scalars
$\phi_u$ and $\phi_v$ with the quantum numbers of the Higgs boson. A
negative mass-squared for one or both Higgs bosons may be introduced
either through a symmetry-breaking plaquette operator at the site
$(1,1)$ \cite{Arkani-Hamed:2001nc} or through the effect of coupling
the Higgs bosons to the top-quark \cite{Arkani-Hamed:2002pa}.  In
either case, the resulting mass-squared of the Higgs is $ |m_h|^2
\simeq {\lambda v^2 \over N^2}~.$

\section{Constraints on Theory Space}

\subsection{Flavor and CP}

Because the light quarks and leptons transform under the $SU(2)\times
U(1)$ gauge interactions at a site in theory
space\cite{Arkani-Hamed:2001nc,Arkani-Hamed:2002pa}, Yukawa couplings
of these fermions to the composite Higgs bosons are generated.  The
FCNC and CP-violation limits derived in Section \ref{sec:flavor} therefore apply.  Because
the composite Higgs bosons are delocalized over the $N^2$ sites of
theory space, the lower bound on $\Lambda$ is a factor of $\sqrt{N}$
stronger.  From D-meson mixing, we have
\beq
\Lambda \gaem 21 \, {\rm TeV} \sqrt{\kappa N\,\left({m_c\over 1.5\,
{\rm GeV}}\right)}~,
\label{eq:Dboundn}
\eeq
so that $\Lambda \gaem \sqrt{N}\cdot 74$ TeV for $\kappa =4\pi$. From CP-violation ($\epsilon$), we have
\beq
\Lambda \gaem 120 \, {\rm TeV} \sqrt{\kappa N\,\left({m_s\over 200\,
{\rm MeV}}\right)}~,
\label{eq:CPboundn}
\eeq
meaning $\Lambda \gaem \sqrt{N}\cdot 425$ TeV for $\kappa =4\pi$.

A significant advantage of theory space models is supposed to be
their ability to produce a light Higgs without fine-tuning.  We must
check how compatable this is with the FCNC and CP-violation
constraints above.

The most important corrections to the Higgs boson masses
arise from the interactions added to give rise to the top-quark
mass. The fermion loop Coleman-Weinberg\cite{Coleman:jx} contribution to the Higgs
mass-squared is of order
\beq
|\delta m^2_H | \simeq {N_c y^2_t M^2 \over 16 \pi^2} \approx 
{N_c y^4_t \over (16 \pi^2)^2}\Lambda^2~,
\label{fermioncw}
\eeq
where $N_c=3$ accounts for color.  In this case, the absence of
fine-tuning ($\delta m^2_H/m^2_H \laem 1$) implies
\beq
\Lambda \laem {16\pi^2 \sqrt{\lambda} v \over \sqrt{N_c} y^2_t N}
\approx {22\,{\rm TeV}\,\sqrt{\lambda}\over N}~.
\label{lambdabound}
\eeq

Comparing eqs. (\ref{lambdabound}) and (\ref{eq:Dboundn}) we see that
remaining consistent with the low-energy constraints makes fine-tuning
inevitable for large $N$.  Even for the smallest $N$, some fine-tuning
will be required.  For example, for $N=2$ ($N = \sqrt{2}$) fine-tuning
on the order of 1\% (3\%) is required by the bound on D-meson mixing.
If the bound from CP violation (\ref{eq:Ebound}) must also be
satisfied, the fine-tuning required is of order .04\% (0.09\%).

\section{Weak Isospin Violation}

The kinetic energy terms for the light composite Higgses
include isospin-violating interactions\cite{Chivukula:2002ww}
\beq
{L}_{kin} \supset - {1\over 6 N f^2} 
\left[(\partial_\mu \phi^\dagger_u \phi_u)^2 
- (\partial_\mu \phi^\dagger_u \phi_u)
(\phi^\dagger_u \partial^\mu\phi_u) +
(\phi^\dagger_u \partial^\mu\phi_u)^2\right] + u \leftrightarrow v~.
\eeq
The resulting contribution to the $\rho$ parameter is\cite{Chivukula:2002ww}
\beq
\Delta \rho^\star  = \alpha \Delta T = {v^2 \over 4 N^2 f^2}
\left( 1- {\sin^2 2\beta \over 2}\right)~.
\label{deltarho}
\eeq
Current limits derived from precision electroweak observables
\cite{Chivukula:1999az} require that $\Delta T \laem 0.5$ at 95\%
confidence level for a Higgs mass less than 500 GeV.  The
bound in eqn. \ref{deltarho} implies that
\beq
\Lambda \simeq 4\pi f \gaem 
{25\,{\rm TeV} \over N} \left(1-{\sin^2 2\beta \over 2}\right)^{1/2}~.
\label{lambdaisobound}
\eeq
Comparison with eq.(\ref{lambdabound}) shows that the underlying
strong dynamics cannot be at energies $\ll$ 10 TeV,
even if the high-energy theory contains approximate flavor and CP
symmetries that avoid the limits of eqs. (\ref{eq:Dbound}, \ref{eq:Ebound}).

\section{Discussion}

Theory space models propose to provide a naturally light composite
Higgs boson without relying on approximate symmetries of the
high-energy underlying strong dynamics.  This talk argues that the
low-energy structure of composite Higgs models does not automatically
make them invulnerable to constraints from FCNC, CP-violation, or
weak-isospin violation.  Assumptions about symmetries of
the underlying dynamics are required (see, e.g., discussion in ref. \cite{Chivukula:2002ww}).

For theory space models based on an $N\times N$ toroidal 
lattice\cite{Arkani-Hamed:2001nc,Arkani-Hamed:2002pa,Arkani-Hamed:2002qx,Arkani-Hamed:2002qy},
the lower limit from FCNC on the scale of strong dynamics is $\Lambda
\geq 74\,{\rm TeV} \sqrt{N}$, implying a minimum bound of 105 TeV.
However, if fine-tuning of the higgs mass is to be avoided
in such models, $\Lambda \leq 22\,{\rm TeV} \sqrt{\lambda}/N$;
preventing FCNC then leads to fine-tuning at the level of
$10/N^3\,$\%.  The lower limits on $\Lambda$ from 
weak isospin violation are weaker than those from FCNC (but
hard to avoid), while those from CP-violation can be much stronger.

\section*{Acknowledgments}
Supported in part by Department of Energy 
under DE-FG02-91ER40676 and by National Science 
Foundation grant PHY-0074274.

\end{document}